\documentclass{WileyMSP-template}

\begin{document}

\pagestyle{fancy}

\title{A simple algorithm for the design of accelerating Bessel beams with adjustable features along their propagation}

\maketitle


\author{Keren Zhalenchuck}
\author{Alon Bahabad*}

\begin{affiliations}
Prof. A. Bahabad\\
Department of Physical Electronics, School of Electrical Engineering, Fleischman Faculty of Engineering, Tel-Aviv University, Tel-Aviv 69978, Israel\\
Tel-Aviv University Center for Light-Matter-Interaction, Tel-Aviv 6997801, Israel\\
Email Address: alonb@eng.tau.ac.il

K. Zhalenchuck\\
Department of Physical Electronics, School of Electrical Engineering, Fleischman Faculty of Engineering, Tel-Aviv University, Tel-Aviv 69978, Israel\\
Tel-Aviv University Center for Light-Matter-Interaction, Tel-Aviv 6997801, Israel\\

\end{affiliations}



\begin{abstract}
We present an extremely simple method for designing self-accelerating non-diffracting beams having arbitrary trajectories while their intensity, width and orbital angular momentum are modulated in a prescribed way along their propagation. Different beams constructed with this method are demonstrated experimentally in the paraxial regime and numerically in the non-paraxial regime. 
\end{abstract}

\section{Introduction}
John H. McLeod introduced the axicon in 1954. It is an optical element that forms a continuous straight focal line, namely a zero-order Bessel beam \cite{McLeod:54}. In 1987 Durning et al. created such a beam using collimated light that illuminates a circular slit located in the focal plane of a lens \cite{PhysRevLett.58.1499}. Bessel beams have drawn a lot of research interest due to their properties: exhibiting  non-diffraction and self healing while possessing an axial symmetric transverse profile \cite{doi:10.1080/0010751042000275259}. Generation of Bessel beams using tunable axicons programmed on spatial light modulators is considered very efficient and is demonstrated and analyzed in a few works \cite{axicon1,axicon2,axicon3}. 

The realization of Airy optical beams in 2007 \cite{Airy} was the first demonstration of spatially self accelerating optical beams. Note that there is a dispute regarding the adequacy of the usage of the term “self-accelerating” though it is quite common for optical beams with a dominant intensity lobe having a curved axial trajectory \cite{doi:10.1119/1.11855, PhysRevA.98.060101}.    

A few works realized Bessel Beam acceleration along various trajectories, e.g. spiraling and snaking \cite{Morris_2010,Jarutis:09,Lee:10,Rosen:95}. Complex methods were developed for the design and observation of Bessel-like beams of arbitrary order that propagate in predesigned trajectories, both paraxial and non-paraxial \cite{bessellike,bessellike_observe,bessellike_OAM,bessellike_Nonparaxial}. Lately, the possibility to control the propagation trajectory of an existing Bessel Beam applying a pure phase modulation was demonstrated \cite{Flexible}.

It was also demonstrated that the polarization state of Bessel beams of arbitrary order, propagating along a straight line,  can be varied along their propagation \cite{Moreno:15}. Moreover, the intensity and hollow core radius of a Bessel beam of arbitrary order were shown to be tunable along the optical axis \cite{Cizmar,tunable_bessel}. Furthermore, Bessel beams with controllable intensity profiles along arbitrary trajectories were designed and observed \cite{Yan:21}. Beams with a varying OAM along their propagation have also been suggested \cite{YangZhuZengLuZhaoCai+2018+677+682,PhysRevA.93.063864,Davis:16}. We treat the OAM as the value of the  number l associated with the presence of the function $e^{i l\phi}$ in the complex amplitude distribution of the beam, where $\phi=atan(\frac{y}{x})$. For beams which are eigenstates of the orbital angular momentum, $l$ is known as the topological charge. The interpretation we use here is a local one, associated with the distribution of the beam around an area of interest.

Here, we present theoretical and experimental results of a  simple and intuitive algorithm for the design of a Bessel beam with adjustable features along an arbitrarily-set propagation trajectory. The set of parameters that can be modified using this algorithm is: the local wave-vector (determining the desired trajectory), the local orbital angular momentum (OAM), the intensity and the beam's width. These parameters are set for a dominant intensity lobe and hence their local nature.  

Potential applications for this type of beams are optical trapping and manipulation  \cite{a881563804b9427d933f0cf730156261,PhysRevLett.75.826,PhysRevA.54.1593,Simpson:97,Hadad,us}, optical communications \cite{OAM_COMM, Dudley:13}, microscopy \cite{microscopy}, corneal surgery \cite{corneal} and material processing \cite{materials,UNGARO2021107398,Dudutis:22}.

\begin{figure*}[hbt!]
\centering
\includegraphics[width=1\linewidth]{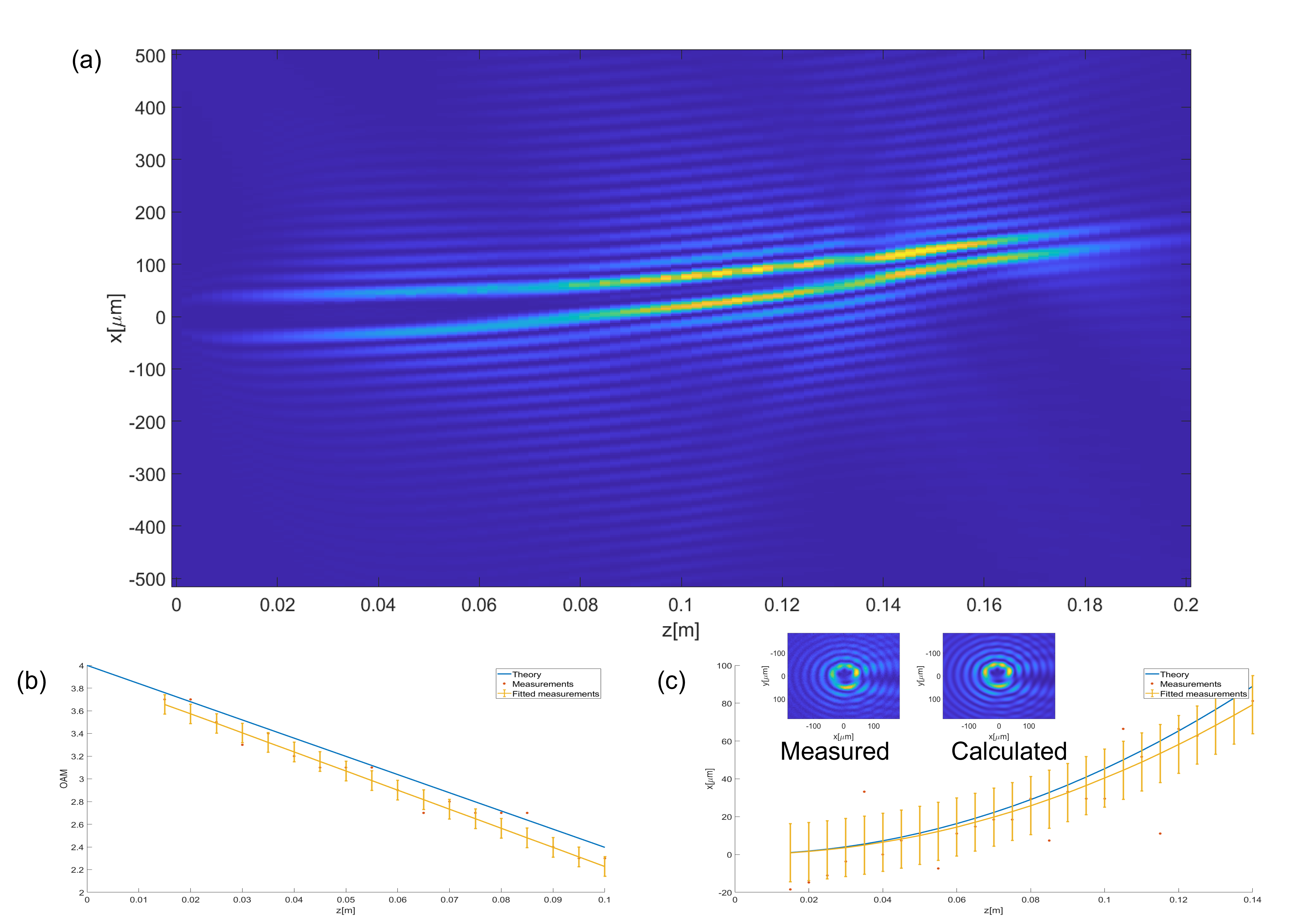}
\caption{\textbf{A curved paraxial Bessel beam with a varying OAM.}
(a) Simulation of a plane cut of a Bessel Beam with a predesigned topological charge. (b) Predesigned topological charge (blue) and experimentally extracted topological charge (red) at different z values along the beam propagation. Fitted curve for the experimental results with its RMSE is shown in yellow. (c) Predesigned beam trajectory (blue), measurements of the center of the beam at different axial locations (red), and a fit of the measurement to a  curve with its RMSE (yellow). Insets in (c) are (right) the beam cross section in simulation  and (left) as captured on a camera at z=2cm.}
\label{fig:fig1}
\end{figure*}

\section{Results}
\subsection{Analytical design}
Before giving a rigorous treatment of our method we provide an intuitive explanation. Consider a collimated beam impinging on an axicon, which is, in effect, a radial diffraction grating.  A ray description of the axicon maps each ray impinging on the axicon  at a given radial position, to an axial position at which it intersects the optical axis. A beam description of the axicon, shows that an impinging Gaussian beam is converted to a Bessel beam. Adding a linear Cartesian phase shift to the axicon, tilts the angle of propagation of the outgoing beam\cite{axicon1}. This can be thought of as changing the angle of the optical axis. Shifting the axicon phase in a transverse direction, shifts laterally the outgoing beam. Combining the beam and ray approach it is apparent that adding a radially varying phase profile to the axicon can divert the rays impinging at every radial position in a prescribed manner. Thus a mapping of rays impinging at a given radius to a local wave-vector, defining the beam trajectory, is established. Further radial phase functionalities can also manipulate the local OAM and beam's width. Adding radial transmission modulation on top of the phase modulation can partially block light directed to specific locations along an arbitrary trajectory, thus allowing to manipulate the intensity along that trajectory.     

\begin{figure*}[t!]
\centering
\includegraphics[width=1\linewidth]{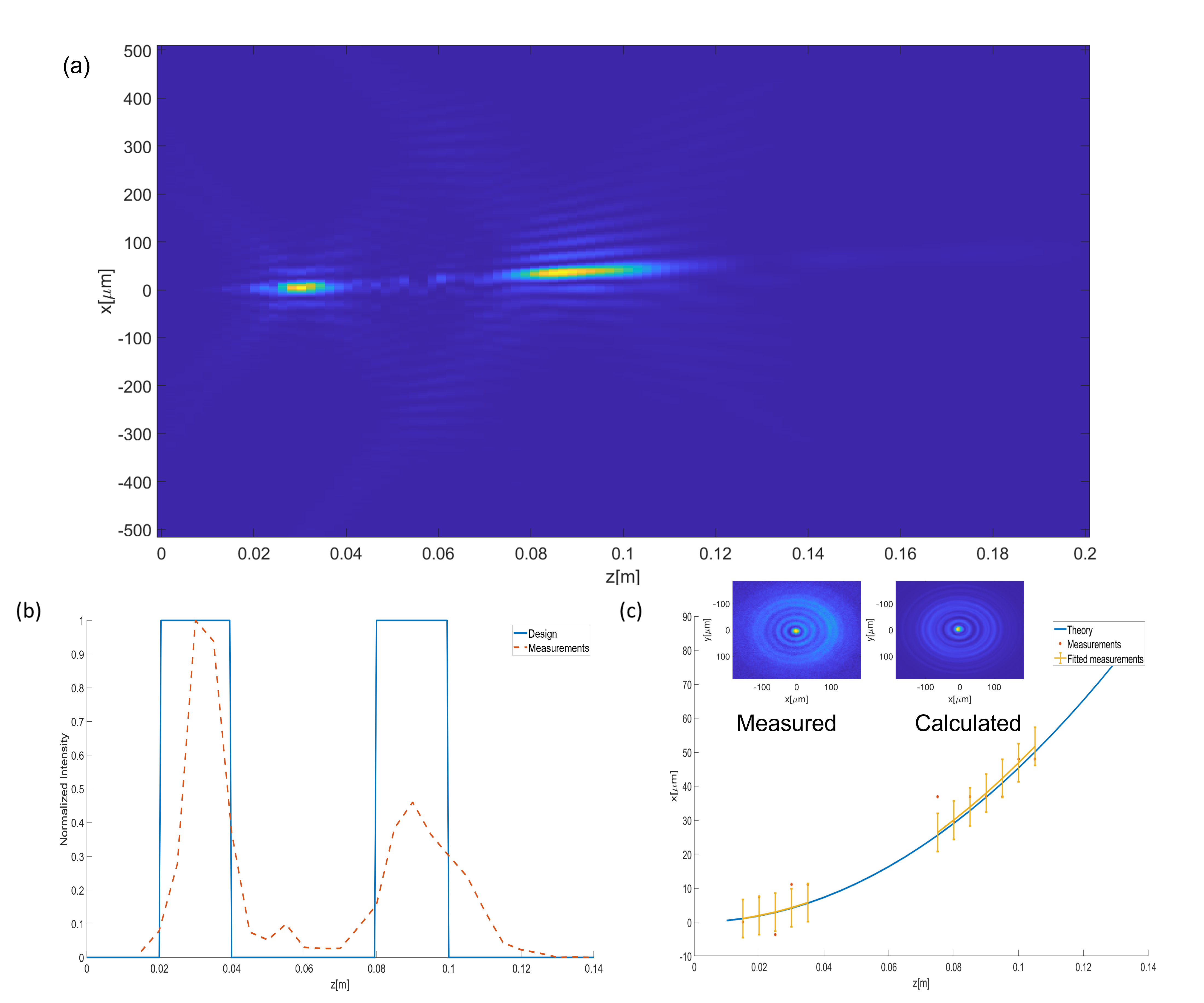}
\caption{\textbf{A paraxial curved Bessel beam with a varying intensity.}
(a) Simulation of a plane cut of a zero order Bessel Beam with a predesigned intensity profile. (b) Predesigned intensity (blue), measured peak intensity normalized to values between 0 to 1 (red). (c) Predesigned beam trajectory (blue), measurements of the center of the beam at the different axial locations where the intensity is higher than 0.1 (red), and a fit of the measurement to a  curve with its RMSE (yellow). Insets in (c) are (right) the beam cross section in simulation  and (left) as captured on a camera at z=2cm.}

\label{fig:fig2}
\end{figure*}
\begin{figure*}[hbt!]
\centering
\includegraphics[width=1\linewidth]{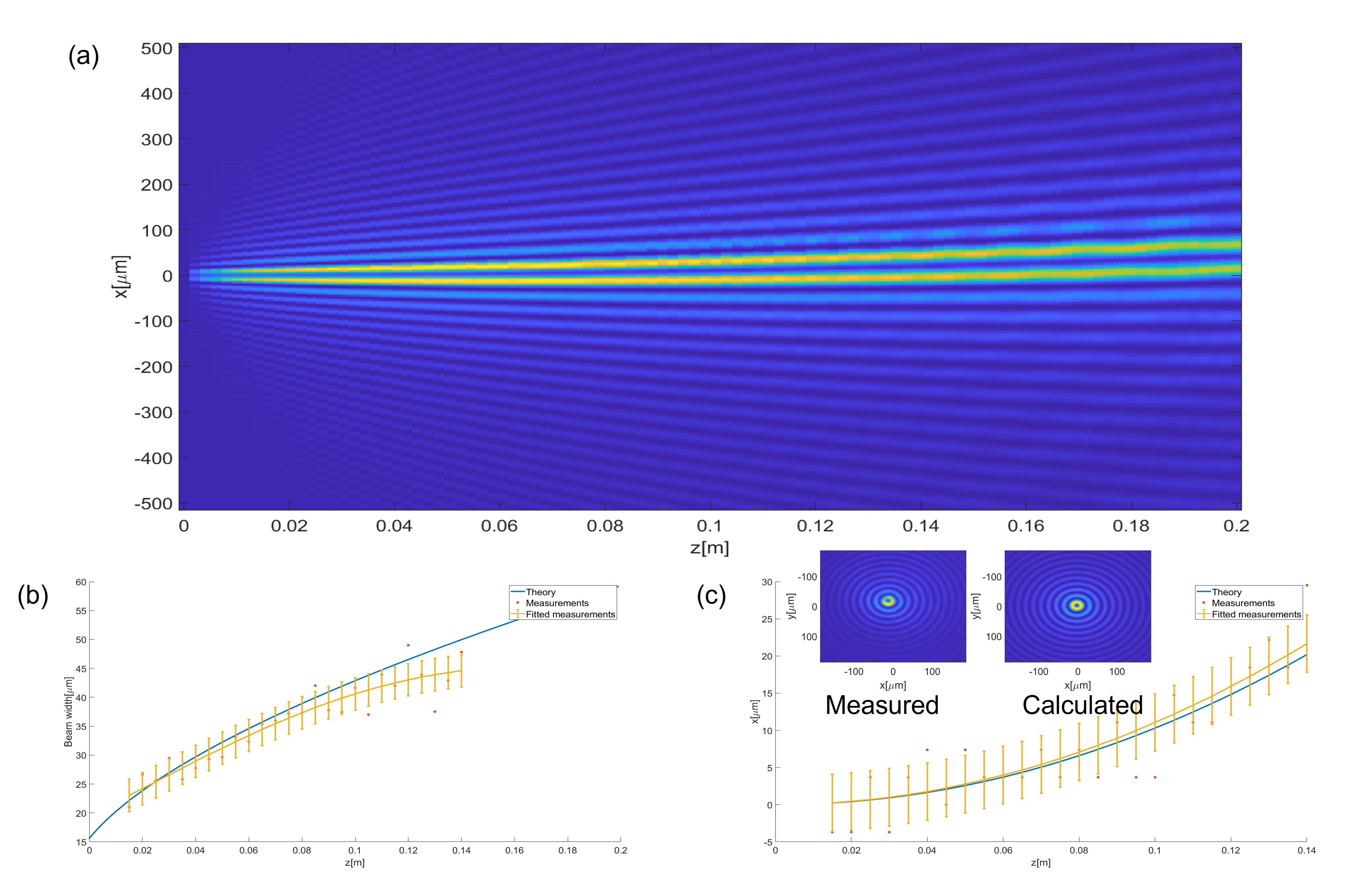}
\caption{\textbf{A paraxial curved Bessel beam with a varying width.}
(a) Simulation of a plane cut of a 1st order Bessel Beam with a predesigned width. (b) Predesigned beam width (blue), and experimentally extracted beam width (red), fitted curve with its RMSE (yellow). (c) Predesigned beam trajectory (blue), measurements of the center of the beam at different axial locations (red), and a fit of the measurement to a curve with its RMSE (yellow). Insets in (c) are (right) the beam cross section in simulation  and (left) as captured on a camera at z=2cm.}
\label{fig:fig3}
\end{figure*}

\subsubsection{Paraxial Regime}
In the paraxial regime we use the Fresnel diffraction integral to evaluate the beams' propagation from an initial field distribution $u_0$ to an output field $u$:
\begin{equation}
u(x, y, z)=  \frac{e^{i k  z}}{i \lambda z}\int_{-\infty}^{+\infty} \int_{-\infty}^{+\infty} u_{0}(\xi, \eta) e ^{\frac{i k}{2 z} [(x-\xi)^{2}+(y-\eta)^{2}]}d\xi d\eta
\label{eq:eq1}
\end{equation}
The transverse coordinates at the observation plane are $(x,y)$ and the transverse coordinates at the input plane are $(\xi, \eta)$  where $r=\sqrt{\xi^{2}+\eta^{2}}$ , $k=\frac{2\pi}{\lambda}$ is the wave-number of the optical beam. The radial diffraction grating equation is:
\begin{equation}
u_{0}(r)=e ^{-i k r sin\theta}
\label{eq:axicon_function}
\end{equation}
Where $\theta$ is the diffraction angle at each radial location $r$, given by the diffraction grating equation $sin\theta=\frac{\lambda}{d}$, where d is the grating period.
We use the stationary phase condition to find the mapping between a ray impinging at $r$ and its crossing point $z$ on the optical axis $(x,y)=(0,0)$:
\begin{equation}
\nabla_{\xi,\eta} \phi_{int} |_{(x,y)=(0,0)} = 0
\label{eq:eq3}
\end{equation}
Where $\phi_{int}$ is the phase of the integrand in eq.\ref{eq:eq1}. The result is:
\begin{equation}
z(r)=\frac{r d}{\lambda}
\label{eq:z_to_r_mapping}
\end{equation}
Knowing the mapping from $r$ to $z$ we can now add more features to the zero order Bessel Beam.
Consider first tilting the rays coming out of the axicon at an angle $\alpha$ with respect to the $x$ axis in the $(x,z)$ plane.  As demonstrated in ref. \cite{axicon1}, this is accomplished by multiplying eq. \ref{eq:axicon_function} by 
\begin{equation}
e^{ik \alpha \xi}
\label{eq:eq5}
\end{equation}
where $\alpha$ is the tilt angle.
This can be verified by noting that now the mapping of eq.\ref{eq:z_to_r_mapping} is the result of the following stationary phase condition: 

\begin{equation}
\nabla_{\xi,\eta} \phi_{int} |_{(x,y)=(\alpha z,0)} = 0,
\label{eq:Stat_phase_cond_tilt}
\end{equation}
establishing that the beam focus is along the line $x=\alpha z$.

It follows that tilting the grating at each radial location by a different angle: 
\begin{equation}
\alpha=\frac{X[z(r)]}{z(r)}
\label{eq:tilt_angle_x}
\end{equation}
 we achieve an arbitrary trajectory $X(z)$ for the main lobe of the beam, under the paraxial approximation.


The same procedure can be expanded to allow also for a varying tilt in the $(y,z)$ plane and thus realize an arbitrary trajectory defined parametrically with $X(z),Y(z)$.

To also get a variation in the local topological charge of the beam along its arbitrary trajectory of  propagation  we can further multiply $u_{0}(r)$ by

\begin{equation}
e^{i l(r(z)) \varphi}
\label{eq:eq7}
\end{equation}
Where $\varphi=arctan(\frac{\eta}{\xi})$ and  $l(r(z))$ is the value of the local topological charge for each $z$ value given by the mapping in eq.\ref{eq:z_to_r_mapping}.

Another tunable parameter is the intensity of the beam. Given the mapping between r to z, modifying the amplitude of $u_0(r(z))$ modifies the beam intensity along the arbitrary trajectory. 

Finally, we would like to also change the width of the beam, Noting that the beam transverse profile is given with the Bessel function $J_{n}(K_{r}(z) r)$, where the transverse local wave vector is $K_{r}(z)=k\frac{r}{z(r)}$. 
The beam width $w$ at each axial location is defined as the distance between the center of the beam and its first transverse zero and is given with:
\begin{equation}
    w=\frac{z_{1n}}{K_r(z)}=z_{1n}\frac{z(r)}{kr},
    \label{eq:Bessel_width}
\end{equation}
where $z_{1n}$ is the first zero of the n-th order Bessel function.  

To allow changing $w$ we let the radial grating period defining the axicon diffraction angle to vary as a function of the radius:  $d=d(r)$. This time applying the stationary phase condition results in the following mapping:

\begin{equation}
    z(r)=\frac{r}{\lambda} \left[\frac{\partial}{\partial r} \left( \frac{r}{d(r)}\right)  \right]^{-1}, \label{eq:z_to_r_2ndmapping}
\end{equation}


Eq.\ref{eq:z_to_r_2ndmapping} constitutes a new mapping, for the case of varying $d(r)$, between the position on the axicon from which a ray is emerging to its crossing point along the axial axis. Note that Eq.\ref{eq:z_to_r_mapping} is a special case of this equation for a constant $d$. This more general mapping allows us to still design arbitrary trajectories and control the intensity and local OAM along that trajectory while supporting a varying beam width.    
Note that although this procedure can be useful, it is not general in the sense that it cannot support any desired form of $w(z)$. Also special care should be given when the $z(r)$ mapping is not one-to-one in a desired domain.  



\begin{figure*}[hbt!]
\centering
\includegraphics[width=1\linewidth]{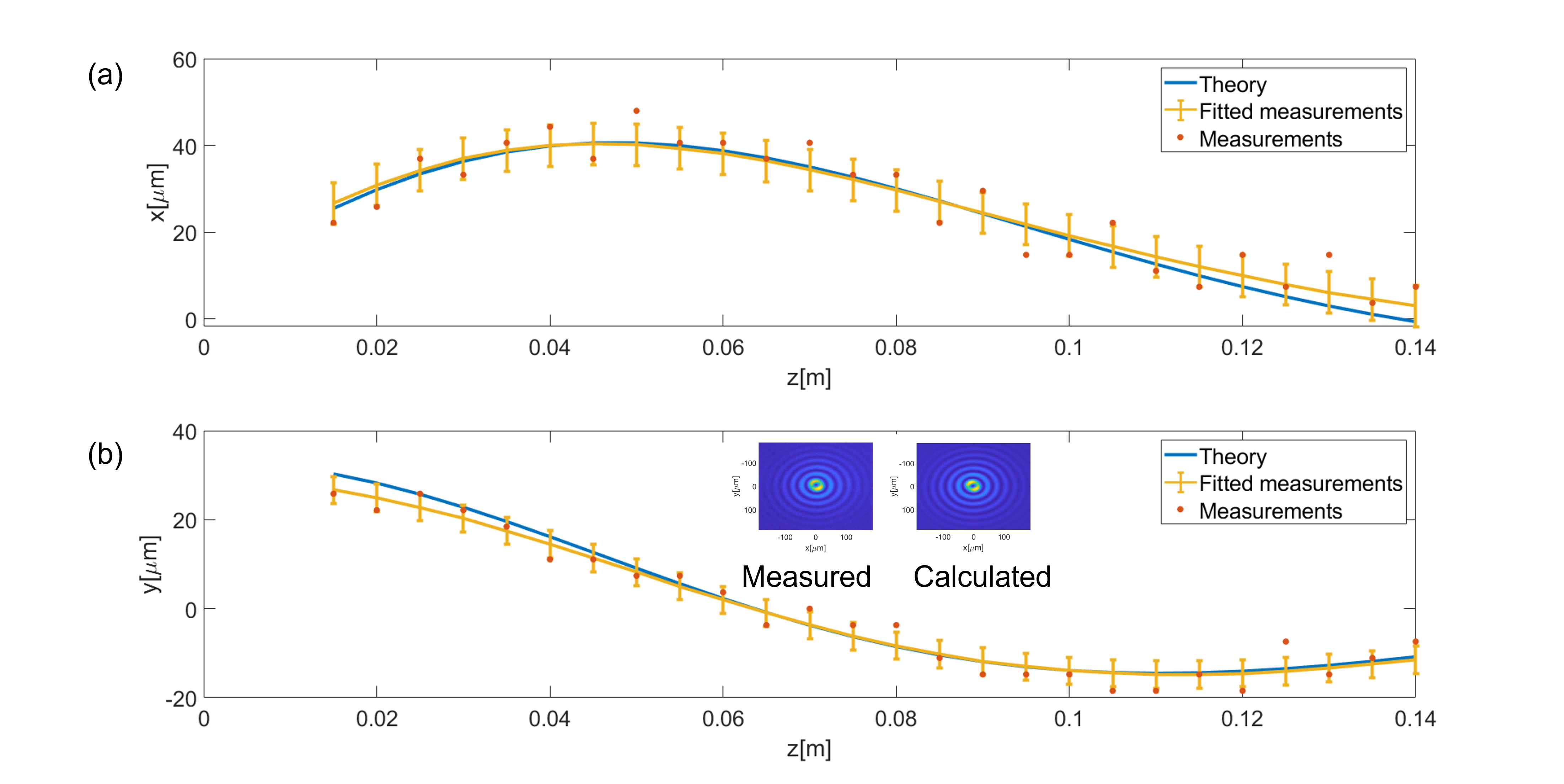}
\caption{\textbf{Curved Bessel Beam with x and y trajectories.}
(a) Predesigned beam x axis trajectory (blue), measurements of the center of the beam at different axial locations (red), and a fit of the measurement to a  curve with its RMSE (yellow).(b)  Predesigned beam y axis trajectory (blue), measurements of the center of the beam at different axial locations (red), and a fit of the measurement to a  curve with its RMSE (yellow).  Insets in (b) are (right) the beam cross section in simulation  and (left) as captured on a camera at z=2cm. }
\label{fig:fig4}
\end{figure*}

\begin{figure*}[hbt!]
\centering
\includegraphics[width=1\linewidth]{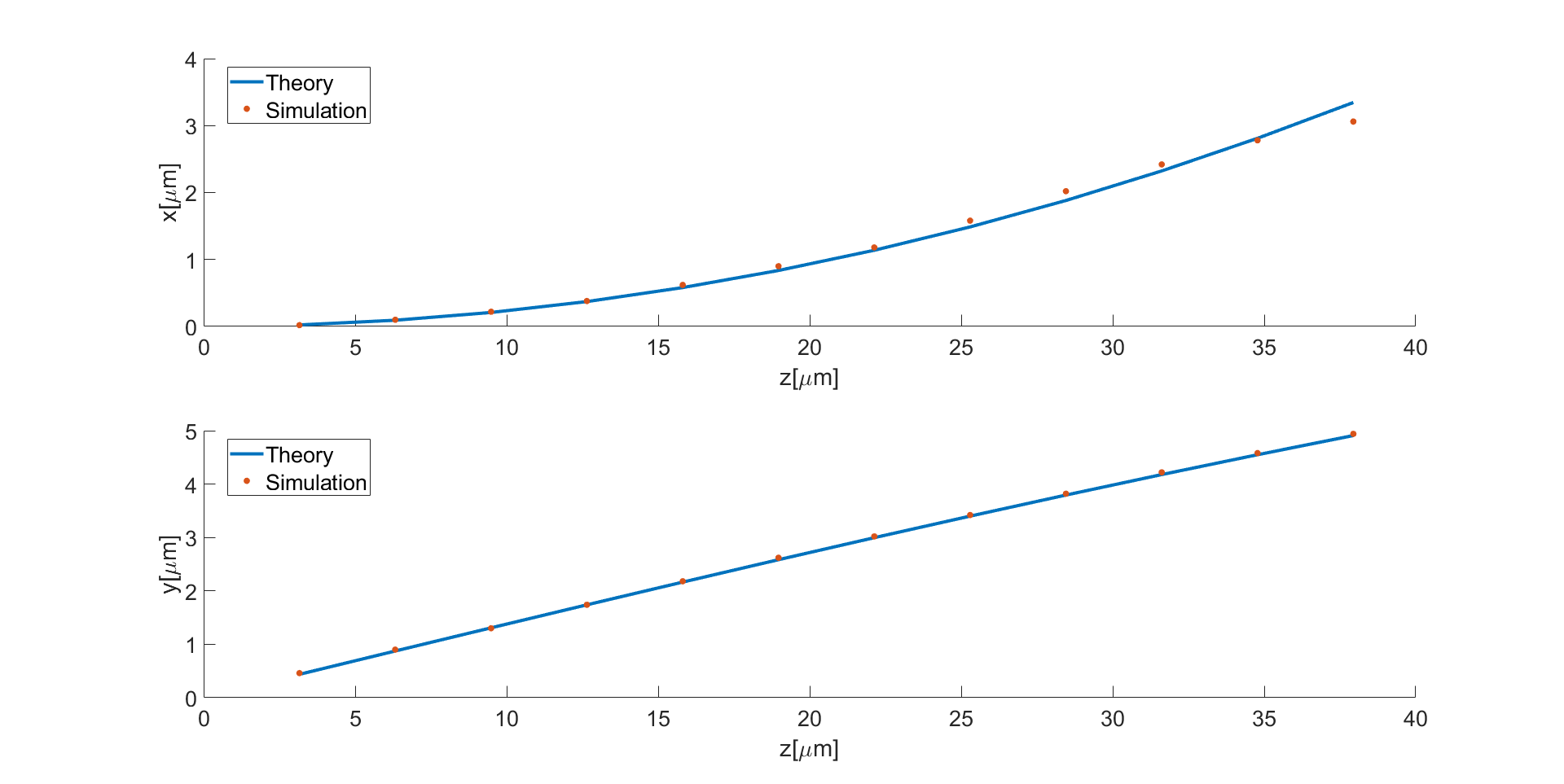}
\caption{\textbf{Curved Non-paraxial Bessel Beam}
(a) Predesigned beam x axis trajectory (blue), simulation of the center of the beam at different axial locations (red).(b)  Predesigned beam y axis trajectory (blue), simulation of the center of the beam at different axial locations (red)).}
\label{fig:fig5}
\end{figure*}

Another method for the generation of beams with predefined trajectories is to move the center of the radial grating, instead of tilting it. This center shift needs to be radially dependent to allow for arbitrary trajectories.  Specifically, the radial coordinates are first mapped, as before, to a specific axial location according to eq.\ref{eq:z_to_r_mapping}, and then the phase function of the axicon is moved to be centered around the coordinates defining the trajectory. Assuming, as before, the trajectory is defined parametrically with $X[z(r], Y[z(r)]$
the axicon phase function is now: 
\begin{equation}
u_{0}(r)=e^{-i k sin\theta r'}
\label{eq:eq12}
\end{equation}
Where $r'=\sqrt{(\xi-X[z(r)])^{2}+(\eta-Y[z(r)])^{2}}$.
Intensity, OAM and width modulations as a function of $z$ can now be added as before.

\begin{figure*}[hbt!]
\centering
\includegraphics[width=1\linewidth]{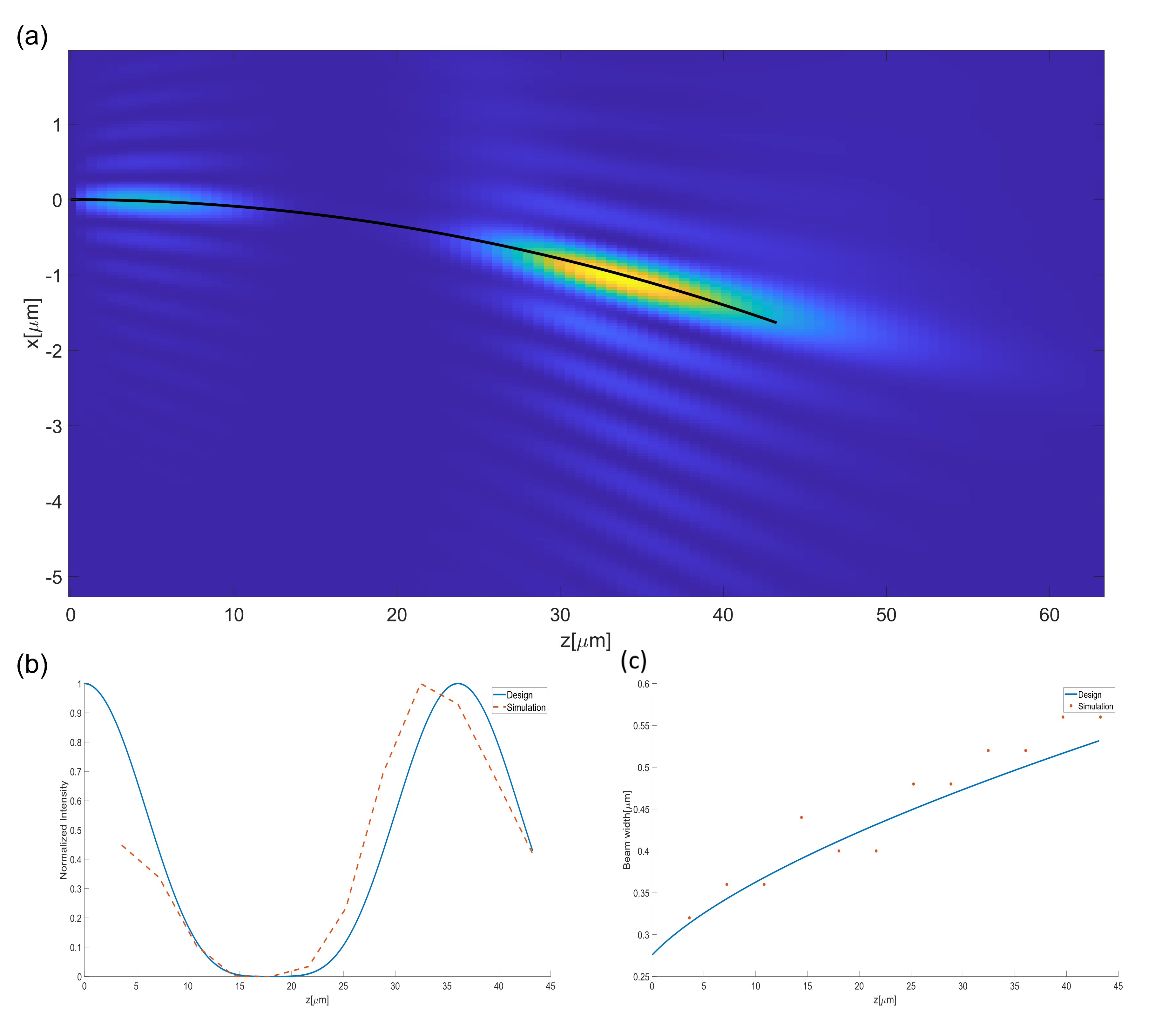}
\caption{\textbf{A Non-paraxial curved Bessel beam with a varying width and intensity.}
(a) Simulation of a plane cut of a zero order Bessel Beam with a predesigned width and intensity profiles and a predesigned beam trajectory (black).(b) Predesigned intensity (blue), measured peak intensity normalized to values between 0 to 1 (red).(c) Predesigned beam width (blue), and simulation extracted beam width (red).}
\label{fig:fig6}
\end{figure*}
{\small
\begin{table}[htbp]
\setlength{\tabcolsep}{1pt}
\centering
\caption{\bf Generated Beams}
\begin{tabular}{ccccc}
\hline
Fig. & modified parameter & Regime & Trajectory [m] & Method \\
\hline
$1$ & Topological charge & paraxial & $X(z)=6.6087\times10^{-4}\times(2.6204z)^2$ & tilt \\
$2$ & Intensity & paraxial & $X(z)=6.6087\times10^{-4}(2.6204z)^2$ & tilt\\
$3$ & Width & paraxial & $X(z)=1.5\times10^{-4}(2.6204z)^2$ & tilt\\
$4$ & None & paraxial & $X(z)=\frac{5.827\times10^{-5}sin(21.6099 z+0.16)}{cosh(13.2746 z+0.16)}$, $Y(z)=\frac{3.3044\times10^{-5} cos(24.697 z)}{cosh(12.3485 z)}$ & shift centers\\
$5$ & None & non-paraxial & $X(z)=8.2609\times10^{-6}(1.6771\times10^{4} z)^2$, $Y(z)=8.2609\times10^{-6}sin(1.6771\times10^{4} z)$  & tilt\\
$6$ & Width, Intensity & non-paraxial & $X(z)=180/\pi\times10^5 ln(cos(\pi z/180))$ & tilt\\
\hline
\end{tabular}
  \label{tab:shape-functions}
\end{table}
}
\begin{figure*}[hbt!]
\centering
\includegraphics[width=1\linewidth]{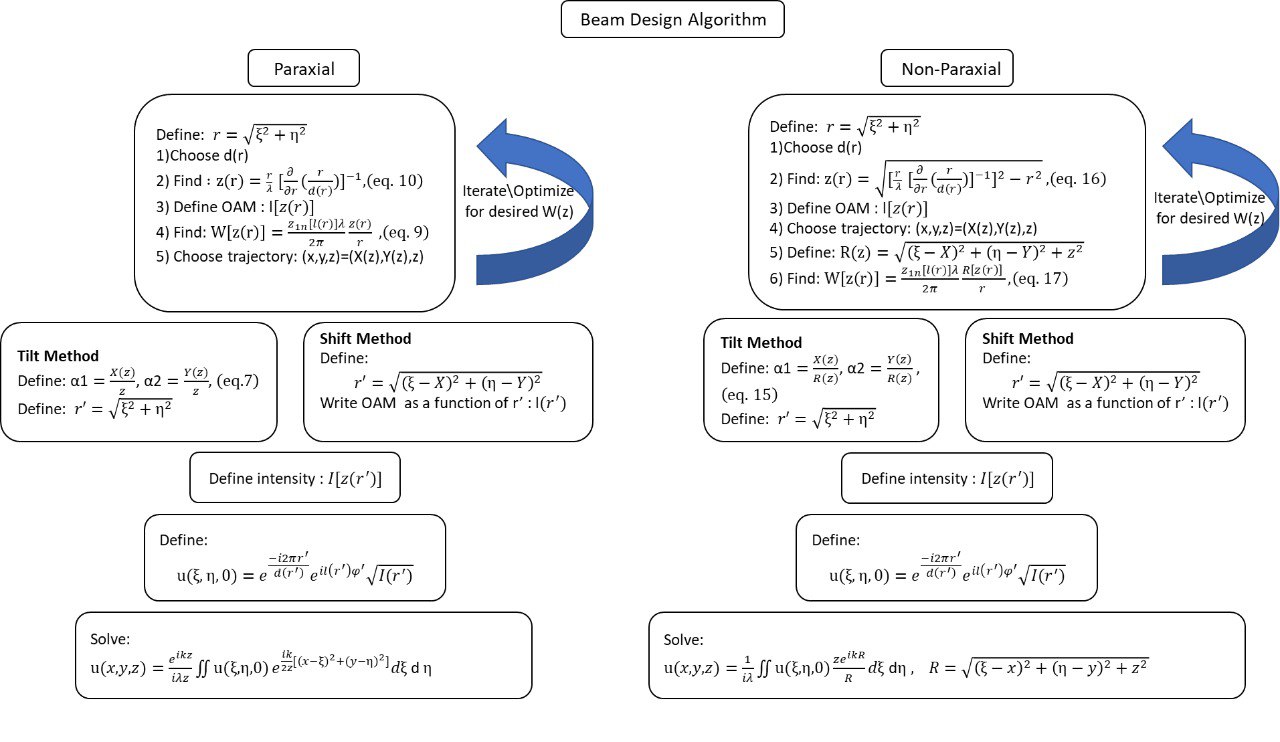}
\caption{\textbf{Beam Design Algorithm}
The algorithm using both tilt and shift methods is presented here for the paraxial and non-paraxial cases.
Here: $l(z), W(z)$ and $I(z)$  are the desired OAM (topological charge), width and intensity of the beam, respectively,  along the axial direction of the beam propagation.}
\label{fig:fig7}
\end{figure*}

\subsubsection{Non-paraxial Regime}
For the non-paraxial case, the phase mask at the input plane has to include features that are smaller than the wavelength and the relevant propagation distance is limited up to several wavelengths \cite{bessellike_Nonparaxial}. In this regime, the Fresnel diffraction integral is invalid and the more general Rayleigh–Sommerfeld diffraction integral is used in order to properly propagate the input field \cite{Rayleigh-Sommerfeld}:
\begin{equation}
u\left(x, y , z\right)=\frac{1}{i \lambda} \int_{-\infty}^{+\infty} \int_{-\infty}^{+\infty} u_0\left(\xi, \eta\right) \frac{z e^{i k R}}{R} \mathrm{~d} \xi \mathrm{d} \eta
\label{eq:sommerfeld}
\end{equation}
Where $R=\sqrt{(x-\xi)^{2}+(y-\eta)^{2}+z^{2}}$.
We use $u_0$ as defined in Eq.\ref{eq:axicon_function} and the stationary phase condition to find the mapping between a ray impinging at $r$ and its crossing point $z$ on the optical axis $(x,y)=(0,0)$ by applying eq.\ref{eq:eq3}, where $\phi_{int}$ is the phase of the integrand in eq.\ref{eq:sommerfeld}. The result is:
\begin{equation}
z(r) = r \sqrt{\frac{d^{2}}{\lambda^{2}}-1}
\label{eq:z_to_r_non_paraxial}
\end{equation}
In order to get a specific beam trajectory $X(z),Y(z)$ by ray tilting, we replace the input field with:
\begin{equation}
u_0(\xi,\eta)=e^{-i k r sin\theta} e^{i k \alpha_{1} \xi}e^{i k \alpha_{2} \eta}.
\label{eq:eq15}
\end{equation}
Using the stationary phase condition:
\begin{equation}
\nabla_{\xi,\eta} \phi_{int} |_{(x,y)=(X(z),Y(z))} = 0,
\label{eq:Stat_phase_cond_tilt_nonparaxial}
\end{equation}
to create a beam focus  along the line $(x,y,z)=(X(z),Y(z),z)$.

we find that: 
$\alpha_{1}=\frac{X(z)}{R}$, $\alpha_{2}=\frac{Y(z)}{R}$ with $R=\sqrt{(X(z)-\xi)^2+(Y(z)-\eta)^2+z^2}$.

Here, similar to the paraxial case, the method of moving the centers of the radial grating can be applied. Assuming, as before, the trajectory is defined parametrically with $X[z(r], Y[z(r)]$
the axicon phase function is again given by eq.\ref{eq:eq12}, where z(r) is given by eq.\ref{eq:z_to_r_non_paraxial}.

Modifying the OAM and intensity along the desired trajectory follows similarly to the paraxial case.

Additionally, the beam width can be modified here similarly to the paraxial case.
The beam width $w$ at each axial location was defined in Eq.\ref{eq:Bessel_width}. 
Similarly to the paraxial case, to change $w$ we let the radial grating period defining the axicon diffraction angle to vary as a function of the radius:  $d=d(r)$. Applying the stationary phase condition in the non-paraxial case results in the following mapping:

\begin{equation}
    z(r)=\sqrt{\left(\frac{r}{\lambda} \left[\frac{\partial}{\partial r} \left( \frac{r}{d(r)}\right)  \right]^{-1} \right)^2-r^2}, \label{eq:z_to_r_2ndmapping_nonparaxial}
\end{equation}

while the width of the beam is given with: 
\begin{equation}
    w(z(r))=\left( \frac{z_{1n} \lambda}{2\pi} \right) \frac{R[z(r)]}{r}. \label{eq:w_vs_r_nonparaxial}
\end{equation}



Eq.\ref{eq:z_to_r_2ndmapping_nonparaxial} constitutes a new mapping, for the case of varying $d(r)$, between the position on the axicon from which a ray is emerging to its crossing point along the axial axis. 
This mapping, which is more general than  Eq.\ref{eq:z_to_r_non_paraxial} allows us to design arbitrary trajectories and control the intensity and local OAM along that trajectory
while supporting a varying beam width in the non-paraxial case. 

A summary of the algorithm presented in this work is shown in Fig.\ref{fig:fig7}.




\section{Experimental Setup}
The experimental setup  consists of a 532 nm CW laser (Laser Quantum Ventus 532 Solo) which reflects off a phase-only spatial light modulator (SLM) (Holoeye Pluto SLM) following expansion and collimation. After the SLM the beam is demagnified with a collimating telescope and imaged using a scientific CMOS camera (Ophir Spiricon SP620U Beam Profiling Camera) on a translation stage set on the optical axis of the beam. The phase profiles for creating the beams were multiplied with a blazed phase grating, directing a background free replica of the beam to the first diffraction order , following the algorithm in \cite{Bolduc:13} for intensity and phase modulation with a single phase-only hologram. The overall phase pattern was loaded to the SLM which was set at one end of a 4f imaging setup. In all the figures that include measured beam trajectories, the positions of the center of the beams were extracted by measuring the difference between the center of a beam  propagating along a straight line and the center of the curved trajectory beams as they were imaged and captured with a camera.

\section{Examples}
We show here several specific beams designed with our simple algorithm. The trajectory of each beam and the parameters that are being modified along the trajectories are indicated in Table.\ref{tab:shape-functions}.  

\begin{enumerate}
  
  \item A paraxial Bessel beam with a varying topological charge using the "tilt" method (see fig. \ref{fig:fig1}). Specifically, the topological charge reduces linearly from 4 to 1 along its propagation. For this we use $l(r)=4-3r_{norm}=4-3\frac{z \lambda}{d m}$  in eq. \ref{eq:eq7} where $r_{norm}=r/m$, with $m$ being a normalization factor such that $r_{norm}=1$ corresponds to the largest radial position available in our lab setup. Experimentally, the topological charge value at each transverse plane along the beam trajectory is extracted following a procedure reported in \cite{unpublished}, the results of which are shown in fig. \ref{fig:fig1}(b).  The planned and measured trajectories agree well with each other as can be seen in fig. \ref{fig:fig1}(c).The error bars indicate RMSE error between the fitted curve (yellow) and the measurements (red).Note that for all the measurements the fit  to a curve was performed using the Non Linear Least Squares method for curve fitting. The equations used for the fit are from table.\ref{tab:shape-functions} with the values of the parameters as initial guesses. Note that the beam is getting narrower with propagation due to the decrease in topological charge as can be seen in fig. \ref{fig:fig1}(a).
  
  \item A curved paraxial Bessel beam with a varying intensity along its propagation using the "tilt" method (see. Fig.\ref{fig:fig2}). Fig.\ref{fig:fig2}(b) shows the designed (in blue) and measured (in red) intensity profiles. The designed intensity profile describing the peak intensity after normalization along the beams propagation is $I(z)=1, 0.02[m]<z<0.04[m], 0.08[m]<z<0.1[m]. I(z)=0, otherwise$. It is apparent that the areas at which the beam intensity is strong is according to the design. The experimental (red) curve was calculated from measurements of the peak intensity of the varying intensity beam at each axial location. This curve was normalized to values ranging from 0 to 1. The trajectory was extracted from measurements where the intensity was higher than 0.1, as seen in fig. \ref{fig:fig2} (b)-(c). 
  \item A curved paraxial Bessel beam with a varying width along its propagation using the "tilt" method (see Fig.\ref{fig:fig3}). The topological charge is kept fixed at a value of $1$. We used a beam with a profile given by a first order Bessel function (in this case $z_{1n}=z_{11}=3.8317$).
  We chose  $d(r)=(8+12 r_{norm})3.2\times10^{-6} [m]$ which, using Eq.\ref{eq:z_to_r_2ndmapping} and Eq.\ref{eq:Bessel_width} yields that the width is given with $w=\frac{3.8317 ((8+12 r_{norm})3.2\times10^{-6})^2}{16 \pi \times 3.2\times10^{-6}} [m]$. Overall the beam width changes from a value of w= 23.05[$\mu$m] at z=1.5[cm] to w=44.58[$\mu$m] at z=14[cm].
  The extracted (red), fitted (yellow) and predesigned (blue) widths are shown in fig. \ref{fig:fig3}(b).
  As seen in fig.\ref{fig:fig3}(a) the beam is getting wider with its propagation along the trajectory which is shown in fig. \ref{fig:fig3}(c).
  

  \item A paraxial Bessel beam curving along a 2D paraemetric curve $X(z),Y(z)$ using the "shift" method (see Fig. \ref{fig:fig4}). The beam has a fixed topological charge of value 1. 
  Note that the previous trajectories were defined with a 1D parametric curve $X(z)$ (that is $Y$ was kept fixed). Additionally while the previous example used the tilt method, employing Eq.\ref{eq:axicon_function}, here we employ the "shift" method, employing Eq.\ref{eq:eq12}.  The trajectory used in this example is shown in fig.\ref{fig:fig4}. It is evident that the measured trajectory agrees well with the designed trajectory. Also in the same figure, the measured  cross section of the beam is almost identical to the calculated one.

  \item A non-paraxial self accelerating Bessel Beam using the "tilt method". The topological charge is fixed at $0$. Here sub-wavelength features are needed and so we  restricted ourselves to computer simulations for this case. 
  As seen in fig. \ref{fig:fig5},
  a non-paraxial x,y trajectory is demonstrated. The simulated (red) and designed (blue) trajectories are in good agreement.

  \item A curved non-paraxial Bessel beam with varying width and intensity along its propagation using the "tilt" method (see Fig.\ref{fig:fig6}). The topological charge is kept fixed at a value of $0$. Fig.\ref{fig:fig6}(b) shows the designed (in blue) and simulated (in red) intensity profiles.The simulation (red) curve was calculated from the peak intensity of the varying intensity beam at each axial location. This curve was normalized to values ranging from 0 to 1. The designed intensity profile describing the normalized peak intensity along the beam propagation is $I(z)=[0.5+0.5cos(3 \pi z \times 1.8491\times10^4)]^2$. We used a beam with a profile given by a zero order Bessel function (in this case $z_{1n}=z_{10}=2.4048$). We chose  $d(r)=(18+8 r_{norm})4\times10^{-8} [m]$ which, using Eq.\ref{eq:z_to_r_2ndmapping_nonparaxial} and Eq.\ref{eq:w_vs_r_nonparaxial} yields that the width is given with $w=\frac{2.4048 \lambda}{2 \pi}\frac{\sqrt{(X[z(r)]-\xi)^2+\eta^2+z[(r)^2}}{r} [m]$. Overall the beam width changes from a value of w= 0.32[$\mu$m] at z=3.6[$\mu$m] to w=0.56[$\mu$m] at z=43.3[$\mu$m]. The extracted (red) and predesigned (blue) widths are shown in fig. \ref{fig:fig6}(c).
\end{enumerate}

 \section{Conclusions}
 We presented a simple method to generate self-accelerating Bessel Beams with  arbitrary trajectories and a predesigned modulation of the topological charge, intensity or width of the beam  along the prescribed trajectory. Any of the above parameters can be varied alone or together with other parameters and thus the variety of different beam types that can be generated is enormous. Our method applies to both the paraxial and non-paraxial regimes. The simplicity of the algorithm and the meager computational resources needed to implement it allow using it in real-time applications where rapid change of either the beam trajectory or any of its parameters is needed.  
 Our results can thus found usage in optical communications \cite{OAM_COMM},optical manipulation \cite{a881563804b9427d933f0cf730156261}, material processing \cite{materials}, quantum optics and imaging \cite{Yao:11}.

\medskip

%
\bibliographystyle{MSP}
\bibliography{sample}

\end{document}